\documentclass[aps,twocolumn]{revtex4}
\usepackage{graphicx}
\usepackage{epsfig}
\begin{document}
\bibliographystyle{apsrev}
\title{Beyond LISA: Exploring Future Gravitational Wave Missions}
\author{Jeff Crowder and Neil J. Cornish}
\affiliation{Department of Physics, Montana State University, Bozeman, MT 59717}

\begin{abstract}

The Advanced Laser Interferometer Antenna (ALIA) and the Big Bang Observer (BBO)
have been proposed as follow on missions to the Laser Interferometer Space Antenna (LISA).
Here we study the capabilities of these observatories, and how they relate to the science
goals of the missions. We find that the Advanced Laser Interferometer Antenna in Stereo (ALIAS), 
our proposed extension to the ALIA mission, will go considerably further toward meeting ALIA's main 
scientific goal of studying intermediate mass black holes. We also compare the capabilities 
of LISA to a related extension of the LISA mission, the Laser Interferometer Space Antenna in 
Stereo (LISAS). Additionally, we find that the initial deployment phase of the BBO would be
sufficient to address the BBO's key scientific goal of detecting the Gravitational Wave Background,
while still providing detailed information about foreground sources.

\end{abstract}

\maketitle

\section{Introduction}

The launch of the Laser Interferometer Space Antenna (LISA)\cite{PrePhaseA} in the next decade
will usher in a new era for gravitational wave detection. LISA will not only detect the presence
of gravitational waves, but also provide detailed information about many thousands of gravitational
wave sources, vastly expanding the burgeoning field of gravitational wave astronomy.

The primary sources for LISA are expected to be compact galactic binaries\cite{H_B_W},
supermassive black hole binaries, and extreme mass ratio inspirals. The gravitational wave
signals detected by LISA encode information about the  parameters of the binary systems,
such as sky locations, masses, and distances, which can be extracted by matched filtering.
Studies into the accuracy with which the source parameters can be recovered from the
LISA data have been performed for compact galactic binaries\cite{Cutler,Hellings,Seto1,CrowderCornish},
supermassive black hole binaries\cite{Cutler,Hellings,Hughes,Seto2,Vecchio}, and extreme mass
ratio inspirals\cite{Barack}.

\begin{figure}
\includegraphics[angle=0,width=0.48\textwidth]{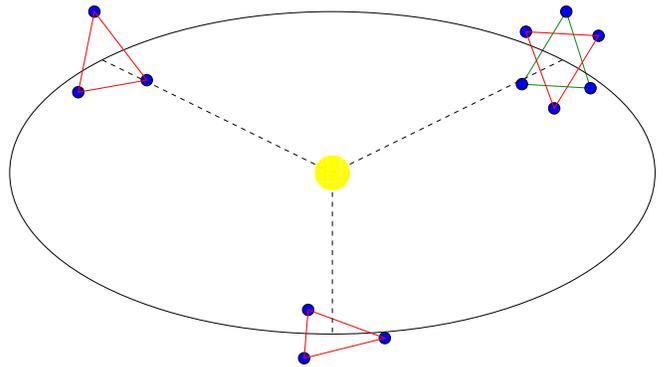}
\caption{\label{BBO_location}The proposed orbital configuration of the Big Bang Observer.}
\end{figure}

It is hoped that LISA will be the first of several efforts to explore the low frequency portion
of the gravitational wave spectrum accessible to space borne interferometers. One possible follow on
 mission is tentatively named the Advanced Laser Interferometer Antenna (ALIA)\cite{ALIA_white_paper},
featuring a spacecraft configuration similar to that of LISA, with smaller arm lengths and
lower noise levels. The primary mission of ALIA will be detecting intermediate mass black holes
(IMBHs), with masses in the range $50 - 50,000 M_\odot$. Information on populations, locations,
and event rates could greatly enhance theories of black hole formation and evolution.
Another proposed follow on mission is the Big Bang Observer\cite{BBO_proposal} (BBO). The BBO will
be an extremely sensitive antenna that is designed to detect the Gravitational Wave Background (GWB)
left by the Big Bang. According to the standard cosmological picture, the GWB is a relic of the early
inflationary period of the Universe. Just as the COsmic Background Explorer (COBE) and the
Wilkinson Microwave Anisotropy Probe (WMAP) missions provided information about the Universe
around the time of last scattering, the BBO should be able to provide information about the
earliest moments in the history of the Universe. The current BBO proposal calls for four LISA-like
spacecraft constellations, with the orbital configuration shown in Figure~\ref{BBO_location}. Two
of the constellations will be centered on a $20^\circ$ Earth-trailing
orbit, rotated $60^\circ$ with respect to each other in the plane of the constellations.  These
constellations will be referred to as the star constellations, as the legs of the constellation sketch
out a six pointed star. The remaining two constellations are to be placed in an Earth-like orbit
$120^\circ$ ahead and behind the star constellations. These two constellations will be referred to
as the outrigger constellations. The purpose of the outrigger
constellations is to provide greater angular resolution for foreground sources (see
Section~\ref{multiple_constellations} for details). The star constellations provide maximum
cross-correlation of gravitational wave signals between the two constellations, with minimal
correlated noise~\cite{Cornish_Larson}. The noise in each detector is expected to be independent,
so over time the overlap of the noise between the two constellations will tend to average to zero, while
the overlap of the signal will grow. The plan is to deploy the BBO in stages, starting with the
star constellation, then adding the outrigger constellations at a later date.

\begin{figure}
\includegraphics[angle=270,width=0.48\textwidth]{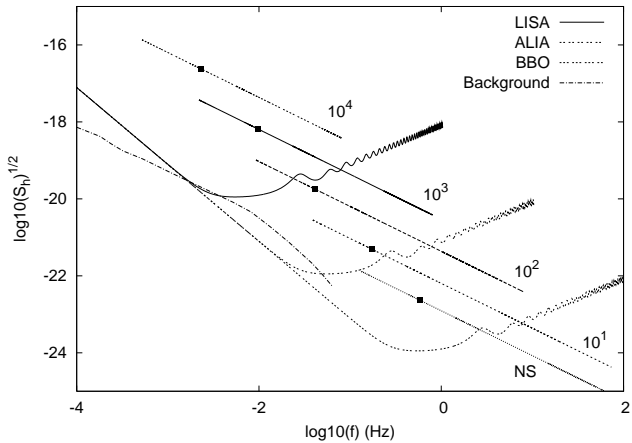}
\caption{\label{char_curves}Sensitivity curves for LISA, ALIA, and BBO with optimally filtered
amplitude plots for equal mass binaries at $z = 1$ in their last year before coalescence. The masses
shown are in units of solar masses. The squares denote the frequency one week from coalescence.}
\end{figure}

Figure~\ref{char_curves} shows the detector sensitivities for LISA, ALIA, and BBO, a level for the expected extra-galactic background confusion noise~\cite{Phinney_Farmer}, and optimally filtered amplitude plots for equal mass binaries in their last year before coalescence
at redshift $z = 1$. The squares shown on the amplitude plots denote the frequency
of the signal one week before coalescence. The signal from coalescing equal mass binaries at
$z = 1$ with masses above $10^1 M_{\odot}$ will be detectable by both ALIA and BBO. The BBO will also be able to detect lower mass systems such as neutron star (NS) binaries. However, for coalescing binaries, a large portion of the signal strength is due to the rapid inspiral in the last week, and for $10 M_{\odot}$ binaries at $z = 1$ or greater, much of this power is deposited
after the signal has crossed the ALIA sensitivity curve.

Here we begin by analyzing the parameter resolution of ALIA for IMBHs. These studies indicate that by
including a second identical constellation of spacecraft in a $20^\circ$ Earth-leading orbit, providing
a constellation separation of $40^\circ$, the parameter resolution will be greatly improved. We will call
this two constellation configuration the Advanced Laser Interferometer Antenna in Stereo (ALIAS). We will
also address the natural question, given the increases in precision that ALIAS provides relative to
ALIA, what would be gained by adding a second constellation to the LISA mission - the Laser Interferometer
Space Antenna in Stereo (LISAS). Lastly, we analyze the parameter resolution of the BBO for NSs,
$10 M_{\odot}$, and $100 M_{\odot}$ black holes (BHs), and for the initial deployment phase of the BBO,
which we call BBO Star, containing the two star constellations of the BBO configuration, but not the
outrigger constellations.
We will show that parameter resolution of the BBO and BBO Star should be sufficient to detect foreground
binary systems, out to and beyond a redshift of $z = 3$.

This paper is organized as follows. We begin with a review of parameter estimation in
Section~\ref{review}. In Section~\ref{multiple_constellations} we describe how a pair
of well separated detectors can significantly improve the angular resolution for transient
sources. This is followed by our study of the ALIA mission in Section~\ref{ALIA_results}, and
a study of our dual-detector variant of the ALIA mission in Section~\ref{ALIAS_results}.
Results for unequal mass binaries for ALIA and ALIAS are in
Section~\ref{log_mass_distribution_results}. The LISA and LISAS missions are compared
in Section~\ref{LISA_LISAS_results}. The BBO mission is studied in Section~\ref{BBO_results},
and results for a down-scoped version of the BBO mission are reported in
Section~\ref{BBO_Star_results}. Concluding remarks are made in Section~\ref{Conclusion}.

\section{Review of Parameter Estimation}\label{review}

In our analysis, the orbits of the binaries are treated as quasi-circular and spin effects are neglected.
This leaves nine parameters that will describe the binary systems: sky location ($\theta, \phi$);
inclination and polarization angles ($\iota, \psi$); reduced and chirp masses ($\mu, \cal M$);
time to coalescence ($t_c$), luminosity distance ($D_L$), and the initial orbital phase ($\gamma_o$).
The signals are modeled using a truncated second-Post Newtonian ($2PN$) approximation\cite{2pn} whereby the
amplitude is kept to Newtonian order while the phase is kept to second order. In other words,
we only include the dominant second harmonic of the orbital frequency.

\begin{figure}
\includegraphics[angle=270,width=0.48\textwidth]{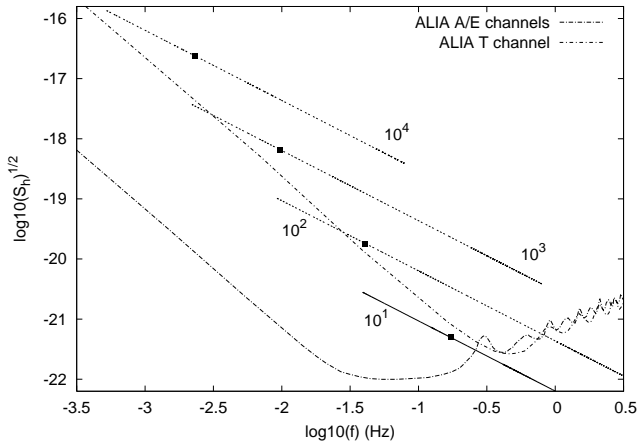}
\caption{\label{AET_ALIA_char_curves}Sensitivity curves for the $A, E$, and $T$ channels for ALIA with optimally filtered
amplitude plots for equal mass binaries at $z = 1$ in their last year before coalescence. The masses shown are
in units of solar masses. The squares denote the frequency one week from coalescence.}
\end{figure}

The response of a space-borne instrument to a gravitational wave source is encoded in the
Michelson-like time-delay interferometry (TDI) variables~\cite{tdi}, $X_i(t)$. Here the
subscript $i$ denotes the vertex at which the signal is read out. In the equal-armlength
limit, the TDI signal, $X_i(t)$, can be formed from a time-delayed combination of
Michelson signals, $M_i(t)$, by
\begin{equation}
X_i(t) = M_i(t) - M_i(t - 2L) \, .
\end{equation}
This differencing cancels the laser phase noise, while preserving the gravitational wave signal.
Rather than work with the correlated $X_i(t)$ variables directly, we use the orthogonal
signal combinations~\cite{PTLA02}
\begin{eqnarray}
A(t) &=& \sqrt{\frac{3}{2}}\left(S_{II}(t) - S_{II}(t-L)\right), \nonumber \\
E(t) &=& \sqrt{\frac{3}{2}}\left(S_{I}(t) - S_{I}(t-L)\right), 
\end{eqnarray}
where
\begin{eqnarray}
S_{I}(t) &=& \frac{1}{3}\left(2 M_1(t) - M_2(t) - M_3(t)\right), \nonumber \\
S_{II}(t) &=& \frac{1}{\sqrt{3}}\left(M_2(t)- M_3(t)\right).
\end{eqnarray}
The $A,E$ combinations cancel the laser phase noise that would otherwise dominate
the Michelson signals. At high frequencies, $f >  f_\ast \equiv c/(2 \pi L)$, 
where $L$ is the length of the detector arms, a third independent data channel $T(t)$ 
becomes available. For frequencies below $f_\ast$ the
$T$-channel is insensitive to gravitational waves and can be used as an instrument
noise monitor. To simplify our analysis we did not use the $T$-channel. Including
this channel would not have had much effect on the trend of our results. As an example, looking at Figure~\ref{AET_ALIA_char_curves} one can see that for ALIA inclusion of the $T$-channel has little to no effect for binaries $< 10^2 M_{\odot}$. Higher mass binaries would have marginally improved signal strength and correspondingly improved resolution of the parameters. Table~\ref{AET_SNR} contains signal-to-noise ratio (SNR) information for the combined $A$ and $E$ channels, as well as the $T$ channel for equal mass binaries as seen by ALIA in the year before coalescence. 
The SNRs in Table~\ref{AET_SNR} are for the final year before coalescence, separated into the contribution from the first $51$ weeks and the contribution from the final week.

\begin{table}[t]
\caption{SNRs for the $T$ channel and the combined SNRs for the $A$ and $E$ channels as seen by ALIA for the equal mass binaries located at $z = 1$.}
\begin{tabular}{|c|cc|cc|} 
\hline
& \multicolumn{2}{|c|}{$51$ Week SNR} & \multicolumn{2}{|c|}{Last Week SNR} \\
 & A+E & T & A+E & T \\ 
\hline
$10^2 M_{\odot}$   &  $44.6$ &  $0.599$ & $16.0$ & $2.29$ \\
$10^3 M_{\odot}$ & $122$ &  $1.23$ & $302$ & $16.1$ \\
$10^4 M_{\odot}$   &  $131$ & $3.52$ & $2145$ & $34.8$ \\
\hline
\end{tabular}
\label{AET_SNR}
\end{table}

In the low
frequency limit, the variables $S_I$ and $S_{II}$ reduce to the independent data channels
defined by Cutler~\cite{Cutler}.  We use the rigid adiabatic
approximation~\cite{rcp} to model the gravitational wave content of the data channels.
The instrument noise in the $A$ and $E$ channels was modeled by~\cite{PTLA02,Nayak}

\begin{eqnarray*}
\lefteqn{S_n(f) = 8\sin^2(f/2 f_\ast) \left[\left(2+\cos(f/f_\ast)\right)S_{\rm pos} \right.} \nonumber \\
& &\left. + 2\left(3+2\cos(f/f_\ast) +\cos(2f/f_\ast)\right)\frac{S_{\rm accl}}{(2\pi f)^4}\right]/(2L)^2\\
\end{eqnarray*}
where $S_{\rm pos}$ and $S_{\rm accl}$ are the one-way position and acceleration noise contributions.
Table~\ref{tab1} lists the instrument parameters used in our study of the LISA, ALIA and BBO missions.
\begin{table}[hbtp]
\caption{Instrument Parameters}
\vspace*{0.1in}
\begin{tabular}{||l|l|l|l||} \hline \hline
Parameter	& LISA	& ALIA	& BBO	\\ \hline \hline
$S_{\rm pos} \; ({\rm m}^2\, {\rm Hz}^{-1})\;$   &  $4\times 10^{-22}$	&  $1\times 10^{-26}$ 
&     $2\times 10^{-34}$     \\ \hline
$S_{\rm accel} \; ({\rm m}^2\, {\rm s}^{-4} \, {\rm Hz}^{-1})\;$ & $9\times 10^{-30}$   
&  $9\times 10^{-32}$     &     $9\times 10^{-34}$   \\ \hline  
$L\; ({\rm m})\;$   &  $5\times 10^{9}$	&  $5\times 10^{8}$ 
&      $5\times 10^{7}$  \\ \hline \hline
\end{tabular}
\label{tab1}
\end{table}

The $A$ and $E$ signals are functions of the nine parameters
$\vec{\lambda} \rightarrow (\theta,\phi, t_c, D_L, \iota, \psi, \cal M, \mu, \gamma_o)$
that describe the source. The source parameters can be recovered from the data
channels using a variety of data analysis methods. We will consider
maximum likelihood estimation, using the log likelihood function
\begin{equation}\label{ml}
\log {\cal L}(\vec{\lambda}') = (s\vert h(\vec{\lambda}')) - \frac{1}{2} 
(h(\vec{\lambda}')\vert h(\vec{\lambda}')) \, ,
\end{equation}
where $s$ denotes the data and $h(\vec{\lambda}')$ denotes the search template.
We use the standard noise-weighted inner product $(a \vert b )$ summed over
the independent data channels:
\begin{equation}\label{nwip}
(a \vert b ) = 2 \sum_{A,E} \int_0^\infty \frac{ a^*(f) b(f) + a(f) b^*(f)}{S_n(f)} \, df\, .
\end{equation}
For our work here the limits of the integration are set by the initial and cut-off frequencies
of the binary systems. To first order these are given by:
\begin{equation}
f_{\rm{initial}} = 3.1 \Big( \frac{10^3 M_{\odot}}{ {\cal M} (1+z)} \Big)^{5/8} \, \rm{mHz},
\end{equation}
\begin{equation}
f_{\rm{cut-off}} = 4.4 \Big( \frac{10^3 M_{\odot}}{M_{total}(1+z)} \Big) \, \rm{Hz}.
\end{equation}
where $M_{total} = {\cal M}^{5/2}/\mu^{3/2}$ is the total binary mass.
Setting $\vec{\lambda}'$ equal to the true source parameters yields the SNR
\begin{equation}
{\rm SNR}^2 = 2 \langle \log {\cal L}(\vec{\lambda}) \rangle =
( h(\vec{\lambda}) \vert h(\vec{\lambda}) ).
\end{equation}
Here the angle brackets $\langle \rangle$ denote an expectation value. The maximum likelihood
estimator $\vec{\lambda}_{\rm ML}$ is defined by:
\begin{equation}
\frac{\partial \log {\cal L}(\vec{\lambda}_{\rm ML})}{\partial \lambda^i} = 0\,.
\end{equation}
The Fisher Information Matrix (FIM), $\Gamma$, is given as the negative of the expectation value
of the Hessian evaluated at maximum likelihood:
\begin{equation}\label{Fisher_equation}
\Gamma_{ij} = - {\Big \langle}
 \frac{\partial^2 \log {\cal L}(\vec{\lambda}_{\rm ML})}{\partial \lambda^i \partial \lambda^j}
{\Big \rangle} = ( h_{,i} \vert h_{, j} ),
\end{equation}
where $h_{,i} \equiv \frac{\partial h}{\partial \lambda^i}$.
For large SNR, the parameter estimation uncertainties, $\Delta \lambda^i$, will have the
Gaussian probability distribution
\begin{equation}
p(\Delta \lambda^i) = \sqrt{\frac{{\rm det} \Gamma}{(2\pi)^9}}\exp\left(-\frac{1}{2}\Gamma_{ij}\Delta \lambda^i \Delta \lambda^j\right) ,
\end{equation}
and variance-covariance matrix
\begin{equation}\label{covariance_equation}
\langle \Delta \lambda^i \Delta \lambda^j \rangle = C^{ij} = \left( \Gamma^{-1} \right)^{ij}.
\end{equation}
The uncertainties in each of the parameters are given by $\Delta \lambda^i = (C^{ii})^{1/2}$
(no summation). One can see from equations (\ref{nwip}) and (\ref{covariance_equation}) that the
parameter uncertainties scale inversely with the SNR.

\section{Multiple Constellation Detection}\label{multiple_constellations}

Detectors such as LISA and ALIA determine the positions of gravitational wave sources through both
amplitude and frequency modulation. The angular resolution improves with time due to the accumulation
of SNR and the synthesis of a long baseline as the detectors move in their orbit. In contrast, a
detector array like the BBO has widely separated elements, and thus has a built-in baseline.
Adding a second widely separated constellation to LISA or ALIA would increase the SNR by a
factor of $\sqrt{2}$, but the main gain in angular resolution for transient sources would be
due to the built in baseline. The advantage of a multi-element array can be understood from the following toy model. Suppose that we have a gravitational wave of known amplitude and frequency. Neglecting the effects of amplitude modulation we have
\begin{equation}\label{toy_signal_equation}
h = A \cos(2 \pi f (t + R \sin(\theta) \cos(2 \pi f_m t + \kappa - \phi)))
\end{equation}
Here $f$ is the source frequency, $R$ is the distance from solar barycenter to the guiding center of
the constellation, $f_m=1/{\rm year}$ is the modulation frequency, $\kappa$ is the azimuthal location
(along Earth's orbit) of the guiding center, and $\theta$ and $\phi$ give the source location on the sky.

With this two parameter ($\theta$ and $\phi$) signal we can analytically derive the uncertainty in
the solid angle for a source observed for a time, $T_{obs}$, by a single constellation
\begin{eqnarray}\label{solid_angle_uncertainty_1_constellation}
\Delta \Omega_{{\rm single}} &=& \frac{2 S_n(f)}{ (A \pi f R)^2 \sin(2 \theta) T_{obs} } \nonumber \\
&& \times \frac{1}{\sqrt{(1-{\rm sinc}^2(2 \pi f_m T_{obs}))}} \, .
\end{eqnarray}
For small observation times $\Delta \Omega_{{\rm single}}$ scales as $T_{obs}^{-2}$, while for
large observation times it scales as $T_{obs}^{-1}$.

Turning to the dual detector case, we simply add together the FIMs for
each individual detector. For two constellations separated by an angle $\Delta \kappa$, this
yields a solid angle uncertainty of
\begin{eqnarray}\label{solid_angle_uncertainty_2_constellation}
\Delta \Omega_{\rm dual} &=& \frac{S_n(f)}{ (A \pi f R)^2 \sin(2 \theta) T_{obs} } \nonumber \\
&& \times \frac{1}{\sqrt{(1-{\rm sinc}^2(2 \pi f_m T_{obs}) \cos^2(\Delta \kappa))}} \, .
\end{eqnarray}
For small observation times and non-zero $\Delta \kappa$, $\Delta \Omega_{\rm dual}$ scales as
$T_{obs}^{-1}$. In other words, the built in baseline leads to a much improved angular resolution for
short observation times. This is very important for coalescing binaries as most of the SNR accumulates
in the final days or weeks prior to merger. Note that if the two constellations are co-located, $\Delta \kappa =0$, the uncertainty in the solid angle is reduced by a factor of two relative to the single detector case by virtue of the increased SNR. Also, note that this toy model only includes Doppler modulation, thus the symmetry between $\Delta \kappa = 0$ and $\Delta \kappa = \pi$. Including the amplitude modulation breaks this symmetry.

\section{Results}

The data shown here is for sources at a redshift of $z = 1$ for ALIA, ALIAS, LISA, and LISAS and redshift
$z = 3$ for BBO, and BBO Star. These correspond to luminosity distances of 6.63 Gpc and 25.8 Gpc,
respectively, using the best fit WMAP cosmology\cite{david}. Each binary system is observed for the
last year before coalescence. Each data point is distilled from $10^5$ random samples of $\theta$, $\phi$, $\iota$, $\psi$, and $\gamma_o$. For $t_c$, $D_L$, $\cal M$, and
$\mu$ we use logarithmic derivatives so that the uncertainties listed have been scaled by the value of
the parameters. The uncertainty in sky location is simply the root of the solid angle uncertainty
($\sqrt{\Delta \Omega}$). The remaining angular parameters, $\iota$, $\psi$, and $\gamma_o$, have not
been scaled.

The angular variables were chosen by using a Monte Carlo method. The values for $\cos(\theta)$ and
$\cos(\iota)$ were chosen from a random draw on $[-1,1]$. Values for $\phi$, $\psi$, and $\gamma_o$ were
each chosen from random draws on $[0,2 \pi]$. The parameters $\cal M$ and $D_L$ were set for each Monte Carlo
run (though they changed between runs). Time to coalescence, $t_c$, was set to $1$ year plus a small offset
so that during the year of observation the binary did not reach a relativistic regime that would not be
properly modeled by the $2\rm{PN}$ approximations used.

The mean SNRs quoted in this paper are calculated by taking the square root of the average of the squares of the individual SNRs. For our analysis, positive detection will be restricted to SNRs above $5$.

\subsection{\label{ALIA_results}Results for ALIA}

Table~\ref{ALIA_histogram_table} summarizes the medians and means of parameter uncertainties
for detections by ALIA of equal mass binaries with masses of $10^2 M_{\odot}$, $10^3 M_{\odot}$,
$10^4 M_{\odot}$, and $10^5 M_{\odot}$. The SNRs for this range of masses shows that ALIA should get
positive detection of $99^+\%$ of coalescing IMBHs located at $z=1$ or closer, which would provide good
information on the coalescence rates. The great precision in the measurement of $\cal M$ and $\mu$ will
provide a clear picture of the constituent masses of the binary systems. Furthermore, the sub-degree
precision in the sky location, combined with luminosity distances known to a few percent, will facilitate
the construction of a three dimensional distribution of IMBHs with which to test theoretical predictions
(see Ref.~\cite{3d} for a related discussion concering LISA and supermassive black holes).

The uncertainties in $t_c$ a month from coalescence (using $11$ months of data) will be on the order
of a few minutes. This will provide warning time for ground-based gravitational wave detectors, as well
as other systems (telescopes, neutrino detectors, etc.) to gather as much and as varied data as possible
about the coalescence.

\begin{figure}
\includegraphics[angle=270,width=0.48\textwidth]{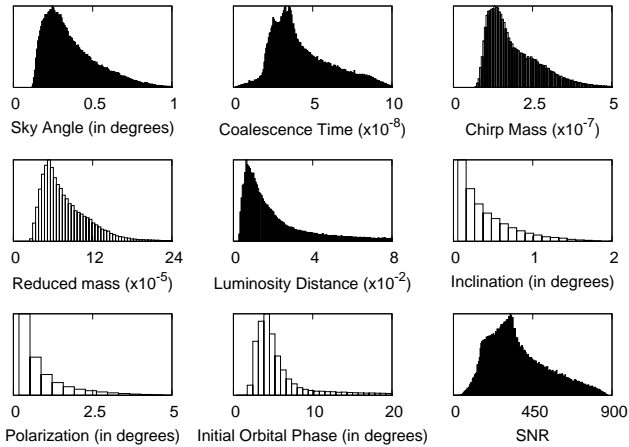}
\caption{\label{ALIA_10_3_histograms} Histograms of the parameter uncertainties and SNR,
for equal mass binaries of $10^3 M_{\odot}$ at $z = 1$, as detected by ALIA.}
\end{figure}

The shapes of the histograms shown in Figure~\ref{ALIA_10_3_histograms}, which are from the
$10^3 M_{\odot}$ data, are representative of the histograms for each of the detectors covered in this work. Note that for $\iota$, $\psi$, and $\gamma_o$, the
tails of the histograms run far beyond the range of the plots shown, raising the values of the means
considerably above their respective median values. For example, while the median value of $\gamma_o$
for equal mass binaries with masses $10^2 M_{\odot}$ is $12.2^\circ$ its mean value is $2185^\circ$,
which is well beyond the $[0,2 \pi]$ range of $\gamma_o$. Uncertainty ranges that are larger than the
possible range of the parameter tell us that the parameter is indeterminate. When uncertainty ranges
exceed the possible range of a parameter, one may drop that parameter from the FIM analysis.
For our analysis, we did not discard any parameters. As an example of how this affects the analysis,
consider the $10^2 M_{\odot}$ study where the mean value of the uncertainties that lie in the
$[0,2 \pi]$ range for $\gamma_o$ would be $36.3^\circ$, while its median would be $9.79^\circ$.
However, for $16.9\%$ of the binaries, the $\gamma_o$ parameter would be indeterminate.

\begin{figure}[b]
\includegraphics[angle=270,width=0.48\textwidth]{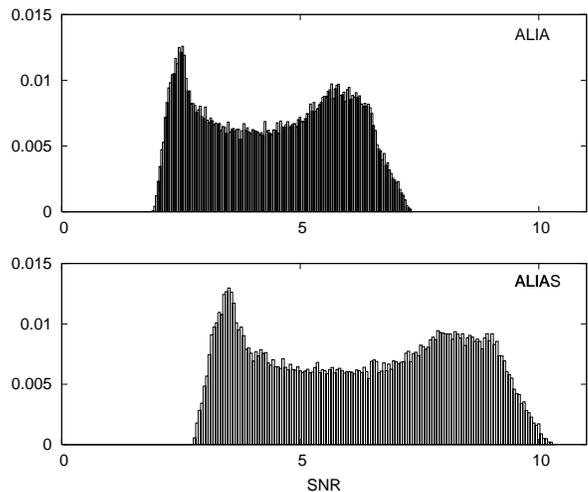}
\caption{\label{ALIA_ALIAS_10_1_sn_histogram} SNR histogram for ALIA and ALIAS from $10 M_{\odot}$
equal mass binaries at $z = 1$.}
\end{figure}

\begin{table*}[t]
\caption{SNR and parameter uncertainties for ALIA with sources at $z = 1$.}
\begin{tabular}{|l|cc|cc|cc|cc|}
\hline
{ }&\multicolumn{2}{|c|}{$10^2 M_{\odot}$}&\multicolumn{2}{|c|}{$10^3 M_{\odot}$}&\multicolumn{2}{|c|}{$10^4 M_{\odot}$}&\multicolumn{2}{|c|}{$10^5 M_{\odot}$}\\
{ }&median&mean&median&mean&median&mean&median&mean\\ 
 \hline
SNR&$71.7$&$76.2$&$466$&$567$&$2937$&$3641$&$6481$&$7970$\\ 
Sky Loc.&$0.169^\circ$&$0.201^\circ$&$0.234^\circ$&$0.264^\circ$&$0.382^\circ$&$0.417^\circ$&$0.518^\circ$&$0.546^\circ$\\ 
$\ln(t_c) \times 10^{-8}$&$1.38$&$1.58$&$2.74$&$3.11$&$4.41$&$5.16$&$5.38$&$6.47$\\ 
$\ln({\cal M}) \times 10^{-8}$&$9.56$&$11.0$&$12.1$&$13.9$&$31.3$&$35.7$&$136$&$152$\\
$\ln(\mu) \times 10^{-5}$&$46.8$&$53.6$&$5.10$&$5.86$&$1.76$&$2.03$&$2.66$&$3.03$\\ 
$\ln(D_L) \times 10^{-2}$&$6.25$&$32.7$&$1.34$&$6.44$&$1.08$&$3.50$&$1.15$&$3.08$\\ 
$\iota$&$3.99^\circ$&$125^\circ$&$0.762^\circ$&$22.4^\circ$&$0.513^\circ$&$8.00^\circ$&$0.515^\circ$&$5.66^\circ$\\ 
$\psi$&$5.44^\circ$&$1092^\circ$&$1.08^\circ$&$176^\circ$&$0.840^\circ$&$42.3^\circ$&$0.931^\circ$&$24.7^\circ$\\ 
$\gamma_o$&$12.2^\circ$&$2185^\circ$&$4.15^\circ$&$352^\circ$&$2.79^\circ$&$85.4^\circ$&$2.51^\circ$&$50.0^\circ$\\ 
\hline 
\end{tabular}
\label{ALIA_histogram_table}
\end{table*}

Figure~\ref{ALIA_ALIAS_10_1_sn_histogram} shows the SNR histograms for $10 M_{\odot}$ equal mass
binaries for ALIA (and ALIAS). As was expected from Figure~\ref{char_curves}, the SNR values
for ALIA are low, with nearly $60\%$ too low for a positive detection. Of note, though, is that the SNR scales roughly as the luminosity distance. This suggests that ALIA should be able to detect nearly all $10 M_{\odot}$ binaries with luminosity distances less than $2$ Gpc. For binaries with masses beyond $15 M_{\odot}$ the low end of the range of SNRs is above $5$, meaning that equal mass binaries with masses in the range of IMBHs should be detectable out to $z = 1$.

\subsection{\label{ALIAS_results}Results for ALIAS}

The main purpose of ALIA is gathering information about IMBHs; our data shows it is capable of doing
this with some success. A more accurate IMBH census could be derived from a dual constellation version
of ALIA that we call ALIA in Stereo or ALIAS.  Each component of the ALIAS constellation
would be offset from the Earth
by $20^\circ$, one in an Earth-trailing and the other in an Earth-leading orbit, giving a $40^\circ$
separation in order to provide increased parameter resolution in the IMBH range.

Table~\ref{ALIAS_histogram_table} summarizes the medians and means of the parameter resolutions
that could be achieved by the ALIAS mission. Results are given for equal mass binaries with
masses of $10^2 M_{\odot}$, $10^3 M_{\odot}$,
$10^4 M_{\odot}$, and $10^5 M_{\odot}$. As expected, the mean SNR increases by $\sqrt{2}$ relative to ALIA. The improvements in parameter resolution are, however, considerably larger.  At the upper end
of the IMBH mass range the angular resolution improves by a factor of $\sim 90$ and the luminosity
distance resolution improves by a factor of $\sim 23$.

\begin{figure}
\includegraphics[angle=270,width=0.48\textwidth]{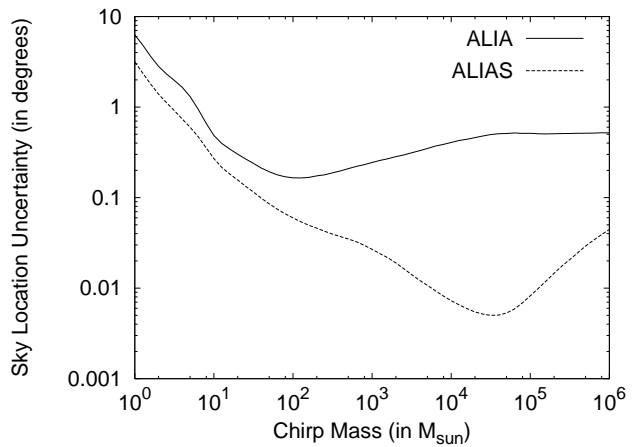}
\caption{\label{ALIA_and_ALIAS_sky_loc_unc_vs_Mc_plot} Uncertainty in the sky location for both ALIA and ALIAS plotted against binary chirp mass. Binary system constituents in this plot have equal masses and are located at $z = 1$.}
\end{figure}

\begin{figure}
\includegraphics[angle=270,width=0.48\textwidth]{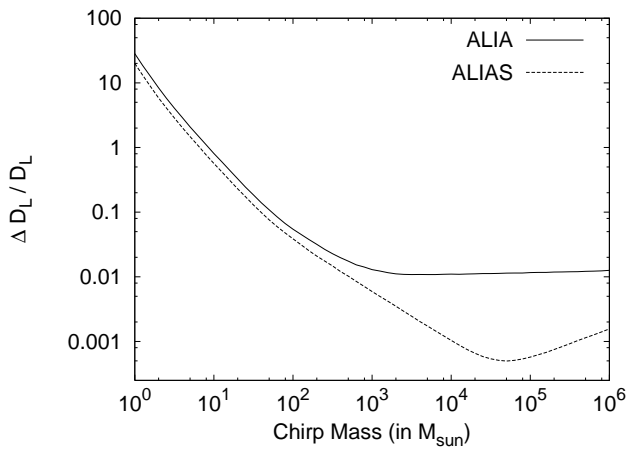}
\caption{\label{ALIA_and_ALIAS_Dl_unc_vs_Mc_plot} Uncertainty in the luminosity distance of both ALIA and ALIAS plotted against binary chirp mass. Binary system constituents in this plot have equal masses and are located at $z = 1$.}
\end{figure}

For masses below
$\sim10^2 M_{\odot}$, ALIAS provides roughly a factor of $\sqrt{2}$ increase in the median values of each of
the parameter uncertainties. This can be seen in the $10^2 M_{\odot}$ data in
Tables~\ref{ALIA_histogram_table} and \ref{ALIAS_histogram_table}, as well as in
Figure~\ref{ALIA_and_ALIAS_sky_loc_unc_vs_Mc_plot} and Figure~\ref{ALIA_and_ALIAS_Dl_unc_vs_Mc_plot},
which plot the uncertainty in the sky location and luminosity distance, respectively, against the chirp mass
of the binaries. The trend shown in these figures holds true for $D_L$, $\iota$, and $\psi$ in that mass range, while the increases in $t_c$, $\mu$, $\gamma_o$, and $\cal M$ are slightly larger, $\sim 1.55$ better than the values for ALIA.
However, the median SNRs are below $5$ for equal mass binaries with chirp masses below $\sim 8 M_{\odot}$
for ALIAS (below $\sim 10 M_{\odot}$ for ALIA).

Figure~\ref{ALIA_and_ALIAS_sky_loc_unc_vs_Mc_plot} also shows that the benefits of a dual constellation
becomes even more significant when the final chirp of the binary occurs above the detector noise
(see Figure~\ref{char_curves}). The main improvement is in the angular resolution, but the decreased
covariances between the sky location and other parameters lead to improved measurements of
$t_c$, $D_L$, $\iota$, and $\psi$ at the upper end of the IMBH mass range. The resolution of these parameters improves (relative to ALIA)
by a factors of $33$, $23$, $18$, and $21$, respectively. The parameters $\cal M$, $\mu$, and
$\gamma_o$ only see a factor of $\sim 2$ improvement in resolution.

Comparing Figure~\ref{ALIA_and_ALIAS_Dl_unc_vs_Mc_plot} with
Figure~\ref{ALIA_and_ALIAS_sky_loc_unc_vs_Mc_plot}, one can see that the increased resolution in the
luminosity distance corresponds with the increased precision in sky location due to their large
covariance. A similar effect was seen in the work of Hughes and Holz~\cite{Hughes_Holz}, where
the addition of electromagnetic information to fix the sky location of a BH merger resulted in
a marked increase in resolution of the luminosity distance. In our case, the degeneracy is broken
by the improved angular resolution afforded by the baseline between the ALIAS detectors.
These improvements in sky location and luminosity distance resolution will provide an even more
detailed three dimensional distribution of the IMBHs than that provided by ALIA. Increasing the
$3D$ resolution of the distribution of IMBHs by roughly a factor of $6$ on the low end of the mass
range ($50 M_{\odot}$) up to $\sim 200,000$ on the high end of the mass range ($50,000 M_{\odot}$), would
provide much more detailed information compared to ALIA, with which to test theoretical predictions
about IMBHs.

\begin{table*}[t]
\caption{SNR and parameter uncertainties for ALIAS with sources at at $z = 1$.}
\begin{tabular}{|l|cc|cc|cc|cc|}
\hline
{ }&\multicolumn{2}{|c|}{$10^2 M_{\odot}$}&\multicolumn{2}{|c|}{$10^3 M_{\odot}$}&\multicolumn{2}{|c|}{$10^4 M_{\odot}$}&\multicolumn{2}{|c|}{$10^5 M_{\odot}$}\\ 
{ }&median&mean&median&mean&median&mean&median&mean\\ 
 \hline
SNR&$102$&$108$&$688$&$802$&$4402$&$5164$&$9630$&$11371$\\ 
Sky Loc.&$0.0620^\circ$&$0.0769^\circ$&$0.0284^\circ$&$0.0330^\circ$&$0.00786^\circ$&$0.00905^\circ$&$0.00757^\circ$&$0.00846^\circ$\\ 
$\ln(t_c) \times10^{-9}$&$7.84$&$9.01$&$10.6$&$11.9$&$3.33$&$4.42$&$2.19$&$2.82$\\ 
$\ln({\cal M}) \times10^{-8}$&$6.04$&$6.86$&$6.64$&$7.61$&$18.3$&$21.1$&$80.8$&$93.0$\\ 
$\ln(\mu) \times10^{-5}$&$23.0$&$26.2$&$2.46$&$2.8$&$1.05$&$1.19$&$1.7$&$1.94$\\ 
$\ln(D_L) \times10^{-3}$&$43.7$&$231$&$6.67$&$34.9$&$1.15$&$5.88$&$0.553$&$2.78$\\
$\iota$&$2.80^\circ$&$89.7^\circ$&$0.413^\circ$&$13.1^\circ$&$0.0678^\circ$&$2.17^\circ$&$0.0324^\circ$&$1.03^\circ$\\ 
$\psi$&$3.86^\circ$&$794^\circ$&$0.575^\circ$&$112^\circ$&$0.105^\circ$&$18.2^\circ$&$0.0501^\circ$&$8.61^\circ$\\ 
$\gamma_o$&$7.96^\circ$&$1589^\circ$&$2.27^\circ$&$226^\circ$&$1.34^\circ$&$37.1^\circ$&$1.23^\circ$&$18.0^\circ$\\ 
\hline 
\end{tabular}
\label{ALIAS_histogram_table}
\end{table*}

Figure~\ref{ALIA_ALIAS_10_1_sn_histogram} shows the SNR histogram for $10 M_{\odot}$ equal mass
binaries for ALIA and ALIAS. While more than two-thirds of the sources at $z = 1$ would be positively detected, nearly one-third would be missed. However, these results suggest that ALIAS will be able to detect nearly all of the $10 M_{\odot}$ binaries with luminosity distances less than $3$ Gpc. This is roughly in keeping with the general trend of ALIAS's capabilities compared to ALIA. 

From Figure~\ref{char_curves} we saw that for masses below $10 M_{\odot}$, the source's signal
lies below the sensitivity curve for the week prior to coalescence. For larger masses,
more of the binary's final chirp will be detectable. This is why the angular resolution of ALIAS becomes
so much better than ALIA's at the upper end of the IMBH mass range. For masses above
$10^5 M_{\odot}$ the final chirp occurs before the ``sweet spot'' of the sensitivity curve,
which diminishes ALIAS's advantage over ALIA.

\subsection{\label{log_mass_distribution_results}Results for ALIA \& ALIAS for unequal mass binaries}

Equal mass binaries are the easiest to study, but they can give an overly optimistic picture of the
instrument capabilities as they yield the smallest parameter uncertainties. To study this bias and
provide a more realistic picture of the capabilities of ALIA and ALIAS, we now consider the case where
the masses of each component in the binary are drawn from logarithmic distributions in the range
$1 - 10^8 M_{\odot}$. The results are still presented as a function of chirp mass, but the mass ratios
reflect the underlying mass distribution.

\begin{figure}
\includegraphics[angle=270,width=0.48\textwidth]{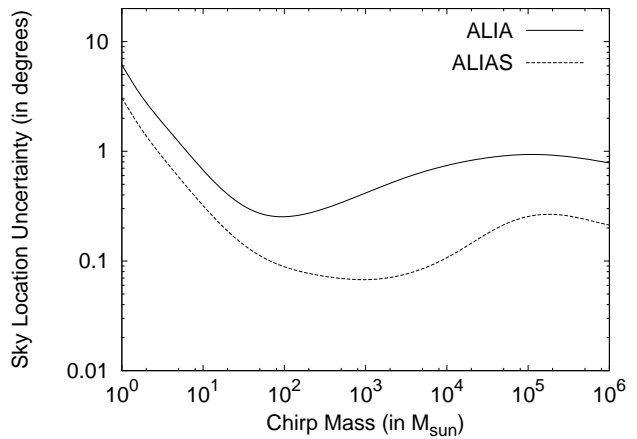}
\caption{\label{lmd__sky_loc_unc_vs_Mc_plot} Uncertainty in the sky location for both ALIA and
ALIAS plotted against binary chirp mass. Binary system constituents in this plot have unequal masses and are located at $z = 1$.}
\end{figure}

Figure~\ref{lmd__sky_loc_unc_vs_Mc_plot} shows the plot of sky location uncertainty against chirp mass.
The uncertainties in sky location for masses above $10^2 M_{\odot}$ are larger than those shown in
Figure~\ref{ALIA_and_ALIAS_sky_loc_unc_vs_Mc_plot}. The increase in sky angle uncertainty relative to
the equal mass case was a factor of $\sim 2$ for ALIA. Similarly, other
parameters show a factor of $\sim3$ increase in their uncertainties relative to the equal mass study.

However, ALIAS still maintains an advantage in locating binaries. In fact, for chirp masses below
$\sim500 M_{\odot}$ the increase in precision for ALIAS over ALIA in locating unequal binaries is
nearly the same as it was for equal mass binaries, a factor of roughly $\sqrt{2}$. At higher chirp masses
we see less of a difference in angular resolution than in the equal mass study (factors of
$\sim 9$ compared to $\sim 90$). This trend holds for
$t_c$, $D_L$, $\iota$, and $\psi$, which show maximum increases in precision $\sim4$ for
ALIAS over ALIA, while $\cal M$, $\mu$, and $\gamma_o$ show maximum increase in precision $\sim2$
for ALIAS over ALIA (as they did for equal mass binaries). Thus the increased resolution in the
$3D$ distribution of IMBHs for ALIAS over ALIA for unequal mass binaries ranges from
a factor of $\sim 6$ up to $\sim 300$.

\subsection{\label{LISA_LISAS_results}Results for LISA and LISAS}

\begin{figure}
\includegraphics[angle=270,width=0.48\textwidth]{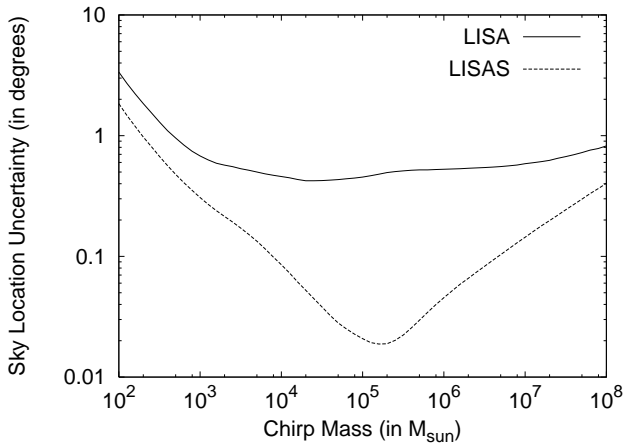}
\caption{\label{LISA_and_LISAS_sky_loc_unc_vs_Mc_plot} Uncertainty in the sky location for both LISA and LISAS plotted against binary chirp mass. Binary system constituents in this plot have equal masses and are located at $z = 1$.}
\end{figure}

Figure~\ref{LISA_and_LISAS_sky_loc_unc_vs_Mc_plot} compares the sky location uncertainties for the
LISA mission to the LISAS mission. As was seen with ALIA and ALIAS, the addition of a second
constellation to LISA provides a marked increase in parameter resolution. The increases in parameter
resolution for LISAS over LISA are similar to those found between ALIAS and ALIA.
The angular resolution showed a maximum improvement of $\sim 25$ just above $10^5 M_{\odot}$.
Also, a lesser benefit than that shown in Figure~\ref{LISA_and_LISAS_sky_loc_unc_vs_Mc_plot} occurs
in the $t_c$, $D_L$, $\iota$, and $\psi$ parameters (with factors of maximum increase $\sim 12$ just
above $10^5 M_{\odot}$), while $\cal M$, $\mu$, and $\gamma_o$ show only a modest maximum improvement by a
factor of $\sim 2$. Our results are in the same range as those found by Seto~\cite{naoki} for sources
detected by LISA that have undergone strong gravitational lensing. In that case the extended baseline was
provided by the time delay in the arrival of the signals, which effectively turned a single
LISA detector into a LISAS system. Seto also gave brief consideration to the performance of a dual
LISA mission, and found the ratio of angular resolution between LISAS and LISA to be larger than
that seen in our simulations. The discrepancy can be traced to Seto including a large galactic confusion
background which limits the time that the sources are in-band, thus magnifying the advantage of the
dual configuration.

\begin{figure}
\includegraphics[angle=270,width=0.48\textwidth]{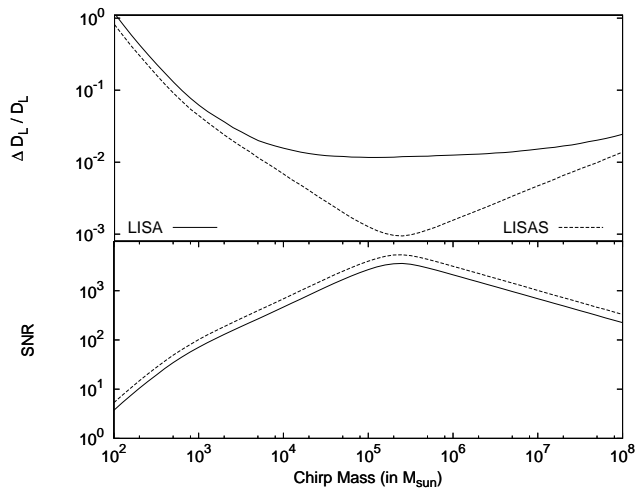}
\caption{\label{LISA_LISA_DL_sn_plot} Uncertainty in the luminosity distance and the SNR for both LISA and LISAS plotted against binary chirp mass. Binary system constituents in this plot have equal masses and are located at $z = 1$.}
\end{figure}

The shift in the placement of the minimum uncertainties, compared to ALIA and ALIAS, can be understood
from Figure~\ref{char_curves}. The signal from a $10^3 M_{\odot}$ equal mass binary is passing below
the LISA sensitivity curve in the last week before coalescence, and before the final chirp of the binary.
This is evident in Figure~\ref{LISA_and_LISAS_sky_loc_unc_vs_Mc_plot}, where the angular resolution
of the single and dual configurations is similar for masses below $10^3 M_{\odot}$.
The sweet-spot of the LISA sensitivity curve is at $\sim 5$ mHz, which is where $10^5 M_{\odot}$ binaries
experience their final chirp. Above $10^5 M_{\odot}$, less of the final chirp is occurs near
the sweet spot, and the difference in the angular resolution diminishes.
Figure~\ref{LISA_LISA_DL_sn_plot} shows how this shift corresponds precisely to
the shift in the maximum SNR of the LISA and LISAS missions. The plot shows how the increase in the mean SNR ($\sqrt{2}$) is uniform across the range of chirp masses. Also, it shows again how
the resolution in $D_L$ improves with the addition of more information in the form of increased
angular resolution.

\subsection{\label{BBO_results}Results for BBO}

Table~\ref{BBO_histogram_table} summarizes the medians and means of the parameter uncertainties
for detections of equal mass binaries with masses of $1.4 M_{\odot}$, $10 M_{\odot}$, and
$10^2 M_{\odot}$ for the BBO. Histograms for this data have the same general shapes as
those shown in Figure~\ref{ALIA_10_3_histograms} and so are not shown.

As can be seen from the data, the BBO is an extremely sensitive detector. The SNRs show that if
the BBO meets design specifications there will be positive detection of coalescing stellar mass
binaries out to (indeed well beyond) a redshift of $z = 3$. Similarly, the BBO will provide a precise
picture of coalescence rates, including how these rates relate to luminosity distance. With a
combination of sky location and luminosity distance the BBO should be able to pick out the host
galaxy of the majority of coalescing binaries. Data taken by the BBO up to a month before
coalescence (using $11$ months of data) will determine $t_c$ to within seconds, again providing
ample warning time for other detectors to gather data on the coalescence.

\begin{table*}[t]
\caption{SNR and parameter uncertainties for BBO with sources at $z = 3$.}
\begin{tabular}{|l|cc|cc|cc|}
\hline
{ }&\multicolumn{2}{|c|}{$1.4 M_{\odot}$}&\multicolumn{2}{|c|}{$10 M_{\odot}$}&\multicolumn{2}{|c|}{$10^2 M_{\odot}$}\\ 
{ }&median&mean&median&mean&median&mean\\ 
 \hline  
SNR&$156$&$162$&$809$&$843$&$4958$&$5189$\\ 
Sky Loc.&$3.31"$&$4.49"$&$0.673"$&$0.920"$&$0.176"$&$0.246"$\\ 
$\ln(t_c) \times10^{-12}$&$179$&$207$&$32.1$&$37.2$&$8.09$&$9.28$\\ 
$\ln({\cal M}) \times10^{-9}$&$63.1$&$72.7$&$7.02$&$8.09$&$30.4$&$35.0$\\ 
$\ln(\mu) \times10^{-6}$&$173$&$200$&$16.3$&$18.8$&$4.97$&$5.70$\\ 
$\ln(D_L) \times10^{-3}$&$28.9$&$150$&$5.58$&$29.1$&$0.927$&$4.73$\\ 
$\iota$&$1.86^\circ$&$57.3^\circ$&$0.360^\circ$&$11.1^\circ$&$0.0592^\circ$&$1.78^\circ$\\ 
$\psi$&$2.54^\circ$&$497^\circ$&$0.492^\circ$&$96.1^\circ$&$0.0806^\circ$&$15.2^\circ$\\ 
$\gamma_o$&$7.71^\circ$&$996^\circ$&$3.93^\circ$&$194^\circ$&$2.82^\circ$&$32.3^\circ$\\ 
\hline
\end{tabular}
\label{BBO_histogram_table}
\end{table*}

\subsection{\label{BBO_Star_results}Results for BBO Star}

As can be seen in Figure~\ref{char_curves}, the final chirp of solar mass or NS coalescing binaries
occurs below the noise level of the BBO. Thus the outrigger constellations do not provide an
extended baseline for the last few days of NS chirp. Positive detection of NS binaries at $z = 3$
can be accomplished with the initial deployment of the two constellations that make up the
star, while still providing the cross-correlating needed to detect the GWB. 

Table~\ref{BBO_star_histogram_table} summarizes the medians and means of the parameter uncertainties
for the detection of equal mass binaries with masses of $1.4 M_{\odot}$, $10 M_{\odot}$, and
$10^2 M_{\odot}$ for the BBO Star. Histograms for this data have the same general shapes as
those shown in Figure~\ref{ALIA_10_3_histograms} and so are not shown.

\begin{figure}
\includegraphics[angle=270,width=0.48\textwidth]{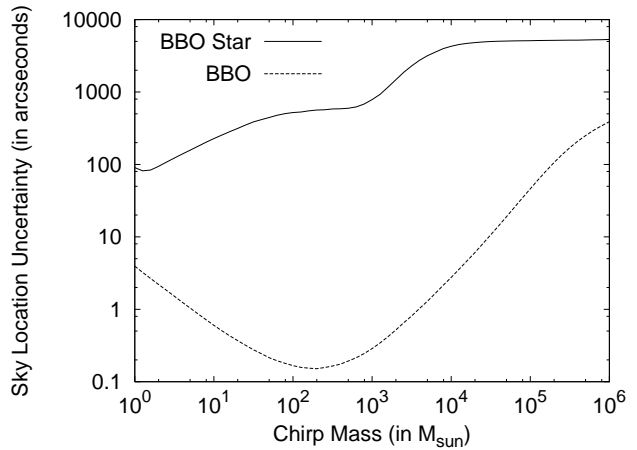}
\caption{\label{BBO_sky_loc_unc_vs_Mc_plot} Uncertainty in sky location versus chirp mass
for the standard BBO design and the star constellations of the BBO. Binary system
constituents in this plot have equal masses and are located at $z = 3$.}
\end{figure}

Figure~\ref{BBO_sky_loc_unc_vs_Mc_plot} plots the uncertainty in sky location against chirp mass for the BBO
and BBO Star. As can be seen, the effect of the outrigger constellations is significant, providing for
increased parameter resolution for the full BBO throughout the range of chirp masses shown, with a maximum
increase of $\sim3700$ around $200 M_{\odot}$. Similar to the ALIA/ALIAS comparison this extra precision
increases until all of the final chirp lies above the sensitivity curve (see Figure~\ref{char_curves}),
and begins to decrease as less and less of the final chirp occurs in the frequency range of the sweet-spot
of the sensitivity curve (which for BBO is $\sim1$ mHz). The improvement in  the mean SNR for BBO over BBO Star is the expected $\sqrt{2}$. Also, the improvement in the resolution of $t_c$ ($\sim 3550$) is comparable
 to that seen in the angular resolution. More modest increases occur for $D_L$, $\iota$, and $\psi$, which increase by factors of $52$, $35$, and $47$, respectively. Only slight ($\sim 2.5$)increases are seen for $\cal M$, $\mu$,
and $\gamma_o$.

The SNRs from BBO Star are sufficient for positive detection of binaries with constituents less than
one solar mass out to, and beyond, $z = 3$. While the full BBO offers considerable adavantages for
doing precision gravitational wave astronomy, BBO Star could fulfill the main science objective of
detecting the cosmic gravitational wave background while still providing useful information about
binary populations.

\begin{table*}[t]
\caption{SNR and parameter uncertainties for BBO star constellations with sources at $z = 3$.}
\begin{tabular}{|l|cc|cc|cc|}
\hline
{ }&\multicolumn{2}{|c|}{$1.4 M_{\odot}$}&\multicolumn{2}{|c|}{$10 M_{\odot}$}&\multicolumn{2}{|c|}{$10^2 M_{\odot}$}\\ 
{ }&median&mean&median&mean&median&mean\\ 
 \hline
SNR&$96.9$&$114$&$484$&$598$&$3022$&$3697$\\ 
Sky Loc.&$82.6"$&$107"$&$214"$&$256"$&$516"$&$596"$\\ 
$\ln(t_c) \times10^{-9}$&$2.98$&$3.43$&$7.52$&$8.60$&$19.8$&$22.3$\\
$\ln({\cal M}) \times10^{-8}$&$13.1$&$15.1$&$2.00$&$2.29$&$6.95$&$7.92$\\ 
$\ln(\mu) \times10^{-5}$&$3.70$&$4.28$&$3.70$&$4.31$&$1.23$&$1.53$\\ 
$\ln(D_L) \times10^{-3}$&$49.9$&$257$&$12.5$&$68.0$&$6.51$&$31.9$\\ 
$\iota$&$3.02^\circ$&$95.8^\circ$&$0.767^\circ$&$22.9^\circ$&$0.310^\circ$&$7.43^\circ$\\ 
$\psi$&$4.07^\circ$&$815^\circ$&$1.04^\circ$&$176^\circ$&$0.529^\circ$&$40.6^\circ$\\ 
$\gamma_o$&$14.1^\circ$&$1633^\circ$&$9.56^\circ$&$357^\circ$&$6.99^\circ$&$84.9^\circ$\\ 
\hline 
\end{tabular}
\label{BBO_star_histogram_table}
\end{table*}

\section{Conclusion/Discussion}\label{Conclusion}

While our survey is by no means comprehensive, it has helped to map out the science that
can be done with the ALIA and BBO missions. We have shown that ALIAS, a modest
extension to the ALIA mission, would be able to return a far more accurate census of the
IMBH population. In addition we have shown that a similar extension to LISA would greatly
improve its ability to locate the host galaxies of coalescing binaries. On the other hand,
if we are willing to give up some of the precision astronomy offered by the full BBO, we
could get by with just the first phase of the BBO deployment. The BBO Star configuration
could satisfy the primary goal of detecting the CGB, while still providing a detailed
binary census.

\begin{acknowledgments}
This work was supported by NASA though the BBO Mission Concept Study led by Sterl Phinney.
\end{acknowledgments}

\end{document}